\documentstyle[a4,12pt,epsf]{article}
\def\ra{\rightarrow}
\def\be{\begin{equation}}
\def\ee{\end{equation}}
\def\bq{\begin{eqnarray}}
\def\eq{\end{eqnarray}}
\begin{document}
\thispagestyle{empty}
\setcounter{page}{0}
\setcounter{page}{0}
\begin{flushright}
WUE-ITP-96-020\\
MPI-PhT/96-108\\
October 1996
\end{flushright}
\vspace*{\fill}
\begin{center}
{\Large\bf 
Sum-Rule Results on Exclusive Decays of Heavy Mesons$^*$}
\\
\vspace{2em}
\large
A. Khodjamirian$^{a,\dagger}$ and R. R\"uckl$^{a,b}$\\
\vspace{2em}
{$^a$ \small Institut f\"ur Theoretische Physik,
Universit\"at W\"urzburg, D-97074 W\"urzburg, 
Germany}\\
{$^b$ \small Max-Planck-Institut f\"ur Physik, Werner-Heisenberg-Institut, 
D-80805 M\"unchen, Germany}\\

\end{center}
\vspace*{\fill}
 
\begin{abstract}
We review QCD sum rule applications to 
hadronic matrix elements of exclusive 
$B$ and $D$ decays. Results are 
presented for 
the form factors $f^+$ and $F_0$ of $B \ra \pi $ and $D\ra \pi$ 
transitions. The predictions are used
to compute the width of $D^* \ra D \pi$, and   
to extract the CKM parameter $V_{ub}$ from the measured $B \ra \pi e \nu$
width. Furthermore, we comment on weak  
annihilation in 
$B \ra \rho \gamma$, as well as on the
radiative decays $D \ra \rho \gamma$,
$D \ra K^* \gamma$, 
and $B \ra \mu \nu \gamma$.
\end{abstract}
 
\vspace*{\fill}
 
\begin{flushleft}

\noindent $^\dagger$ {\small on leave from 
Yerevan Physics Institute, 375036 Yerevan, Armenia } \\

\noindent $^*${\it talk presented by A. Khodjamirian at 
28th International Conference on High Energy Physics, Warsaw, 
July 1996 }
\baselineskip=16pt                                 
\end{flushleft}
 
\newpage

\section{Introduction}

At this conference, a large amount of new experimental data on  
exclusive decay modes of heavy hadrons 
have been presented. In order
to interprete these measurements 
in the framework of the standard model,  
one has to be able to calculate hadronic matrix elements of products 
of quark/gluon operators. Since these matrix elements  
involve long-distance 
quark-gluon dynamics, nonperturbative methods are needed.
In this talk, we report on recent progress in applying 
QCD sum rule \cite{SVZ} methods to solve this problem. 
The basic tools 
are the operator product expansion (OPE)
near the light-cone and dispersion relations in
combination with quark-hadron duality. 
The basic nonperturbative input are the light-cone wave functions 
of light mesons. The wave functions of $\pi$ and $K$ mesons are known
\cite{exclusive} with sufficient accuracy to allow for reliable
predictions 
of heavy-to-light form factors and couplings.
Recently, one has
also 
begun to study applications requiring $\rho$--meson \cite{exclusive} 
and photon \cite{BBK} 
wave functions. 

After a brief outline of the calculational procedure (section 2), 
results are presented for the $B\ra \pi$ and  
$D \ra \pi$ form factors (section 3), and the hadronic
$B^*B\pi$ and $D^*D\pi$ couplings (section 4). From that
we predict the widths of
$D^*\ra D\pi$ and $ B \ra \pi l \nu $. 
The latter is used to extract 
the value of $V_{ub}$ from the CLEO measurement of 
$ B \ra \pi l \nu $ (section 5). Finally, we demonstrate 
the flexibility of light-cone sum rules by summarizing   
results on 
the radiative decays
$B \ra \rho \gamma$,  
$D \ra \rho \gamma$, $D \ra K^* \gamma$,  and $B \ra \mu \nu \gamma$.

\section{Outline of the light-cone sum rule approach}
For definiteness, let us focus on the transition
matrix element 
$\langle L \mid \bar{q}\Gamma_a Q \mid H \rangle $ 
between a heavy ($H$) and a light ($L$) meson. 
The basic object to be calculated is the  
correlation function
\be
F_{ab}(q,p) = \int d^4x e^{ip\cdot x}
\langle L(q) \mid T\{\bar{q}(x)\Gamma_a Q(x), 
 \bar{Q}(0)\Gamma_b q'(0)  \} \mid 0 \rangle~. 
\label{corr}
\ee
Here, $L$ is taken on-shell with momentum q 
and the quark current 
$\bar{Q}\Gamma_b q' $ with the external momentum
$(p+q)$ is chosen 
to carry the quantum numbers of $H$. The appropriate combinations
of Dirac matrices are denoted by 
$\Gamma_{a,b}$. 

In contrast to conventional sum rules \cite{SVZ} employing the Wilson OPE
in terms of local operators, here  
the $T$-product of currents in (\ref{corr}) is expanded 
in terms of nonlocal operators near the light-cone, i.e. at $x^2 \simeq 0$. 
Schematically, this expansion has the form:
\be
\langle L(q) \mid T\{\bar{q}(x)\Gamma_a Q(x), 
\bar{Q}(0)\Gamma_b q'(0)  \} \mid 0 \rangle
\sim \sum_i C_i(x) \langle L(q) 
\mid O_i(x,0)
\mid 0 \rangle 
\label{matrix}
\ee
with calculable coefficients $C_i(x)$. The 
bilocal operators $O_i(x,0)$  are constructed out of 
quark and gluon fields. 

Furthermore, the matrix elements of $O_i(x,0)$ appearing 
in (\ref{matrix})   
are parametrized in terms of so-called 
light-cone wave functions $\phi_i^n$
characterized by 
particle multiplicity and twist. For the leading two-quark operator 
$O_{2a}(x,0)= \bar{q}(x)\Gamma_a q'(0)$ 
one has, schematically, 
\be
\langle L(q) \mid O_{2a}(x,0) \mid 0 \rangle 
=\sum_n \int_0^1 du e^{iuq\cdot x }\phi_{2a}^n(u)(x^2)^n~,
\label{wf}
\ee  
where a path-ordered gauge factor has beed omitted for brevity.
In the momentum region $(p+q)^2 \ll m_Q^2$, the light-cone
expansion is dominated by 
the lowest-twist wave functions, and can therefore be truncated
after a few terms. Thereby, one may keep 
the momentum transfer $p^2$ timelike, in the range
$0 < p^2 < m_Q^2 -O(1 $GeV$^2)$. This 
is very important since it allows to calculate form factors 
in a wide kinematical 
region without the need of extrapolations.

The connection of the correlation function $F_{ab}$ 
with the matrix element 
$\langle L\mid \bar{q} \Gamma_a Q \mid H\rangle$
of interest is obtained by writing a dispersion relation 
in the variable $(p+q)^2$ for $F_{ab}$ and inserting 
the complete set of 
intermediate states $\mid i \rangle $ with the quantum numbers 
of $H$:  
\be
F_{ab}=\sum_i
\frac{
\langle L\mid \bar{q} \Gamma_a Q \mid i\rangle
\langle i \mid  \bar{Q}\Gamma_b q' \mid 0 \rangle
}{m_i^2-(p+q)^2}~.
\label{hadr}
\ee
Obviously, the desired matrix element is contained in 
the contribution from $\mid i \rangle = \mid H \rangle $.   
In order 
to determine it, one has to subtract the contributions of all other
states $\mid i \rangle \neq \mid H \rangle $. For the ground state
$H$ in a given channel, this is indeed possible to a reasonable 
approximation. To this end, one invokes    
quark-hadron duality to estimate the contribution 
of the heavier states in (\ref{hadr}), 
and applies the Borel transformation
in order to exponentially damp their contribution, and to diminish 
the sensitivity of the resulting expression for 
$\langle L\mid \bar{q} \Gamma_a Q \mid H\rangle $
to the duality approximation.

One of the principal advantages of the QCD sum rule method  is the
universality of the 
nonperturbative input parameters,
in the light-cone variant the nonasymptotic 
coefficients of the light-cone wave functions.
Once these parameters have been determined from one set 
of measurements, one can apply the method to other 
observables without having to introduce new unknown parameters.  
Furthermore, the matrix element 
$\langle H \mid  \bar{Q}\Gamma_b q' \mid 0 \rangle$, 
multiplying the $H \ra L$ transition element in (\ref{hadr}),
and also the threshold parameter $s_0$, separating in $(p+q)^2$ the 
ground state at $(p+q)^2= m_H^2$ from the region $(p+q)^2 > s_0$ 
included in the duality integral, can be estimated separately
from appropriate two-point sum rules. 

Another principal advantage 
is the possibility to estimate the accuracy of a given result within 
the same sum-rule framework. An analogous estimate is not 
possible, for example, in phenomenological
quark models. Currently, one main source of uncertainties are 
perturbative corrections that are not yet known for sum rules 
of form factors and hadronic couplings. This problem will 
certainly be solved in near future. More difficult is it to improve 
our knowledge of the nonasymptotic terms  
in the light-cone wave functions. 
The lack of complete understanding leads to about 15-20\% uncertainty in 
the predictions presented below.

Last but not least, 
QCD sum rules automatically include effects due to the finiteness 
of the physical quark masses and can deal with heavy and light quarks.
Moreover, they are not or at least less restricted to particular kinematical 
conditions such as zero recoil in form factors. In this
respect, the sum rule method is not challenged by HQET (heavy quark
effective theory) and complements lattice calculations. 

\section{The $B \ra \pi$ and $D \ra \pi$ form factors}

As a first example, we consider
the matrix element of the $B\ra \pi$ transition
which is parametrized by two form factors:
\be
\langle\pi(q)\mid \bar{u}\gamma_\mu b\mid B(p+q)\rangle
=2f^+(p^2) q_\mu +(f^+(p^2)+f^-(p^2)) p_\mu ~.
\label{matr}
\ee
In  \cite{BKR}, the light-cone
sum rule was obtained for the form factor $f^+$ in the
region of momentum transfer $p^2 < m_b^2-O($ GeV$^2)$,
and to twist 4 accuracy. 
Recently \cite{KRW}, the calculation 
of (\ref{matr}) has been completed by deriving 
the sum rule for $(f^+ + f^-)$ to the same
accuracy.
With these results at hand, we can also 
predict the scalar form factor 
\be
F_0(p^2) = f^+(p^2) +\frac{p^2}{m_B^2-m_\pi^2}f^-(p^2) ~.
\label{f0}
\ee

In Fig. 1, we show our results \cite{BKR,KRW} 
for $f^+$  and $F_0$. 
In order to estimate the form factor $f^+$ at larger $p^2$, 
one may use the single pole approximation involving the $B^*$
resonance. 
The residue of the pole is proportional to the 
$B^*B\pi$ coupling which can be extracted from 
the same correlation function (\ref{corr}) used also 
for $f^+$ (see also section 4). 
As can be seen in Fig. 1, the pole approximation normalized in this
way 
nicely matches 
the result of the direct calculation at intermediate values of  $p^2$. 
Also interesting to note is the consistency of our result on  
$F_0$ with 
the Callan-Treiman relation  
$lim_{p^2 \ra m_B^2}F_0(p^2) = f_B/f_\pi$ \cite{Vol}
although this constraint is weak because of the sizeable 
uncertainty in the $B$ decay constant. 

The corresponding $D \ra \pi$ form factor $f^+$ 
is shown in  Fig. 2.  
Interestingly, in this case there is almost no difference between $F_0$ 
and $f^+$.

\section{The $D^* \ra D\pi$ decay width}

As mentioned before, the correlation
function (\ref{corr}) can also be used to calculate 
the $B^*B\pi$ and $D^*D\pi$ couplings \cite{BBKR}.
The former is defined by 
\be
\langle B^{*0}(p)\pi^+(q)\mid B^+(p+q)\rangle =
-g_{B^*B\pi}q_\mu \epsilon ^\mu . 
\label{g}
\ee
In this application, one has to employ dispersion relations and 
sum rule methods  simultaneously in 
the $B$- and $B^*$- channels. 
After double Borel transformation  one obtains an expression for 
$g_{B^*B\pi}$ the leading twist term of which 
depends on the pion wave functions at the momentum fraction
$ u \simeq 1/2 $. Numerically, we find 
$g_{B^*B\pi}= 29\pm 3 $, and by an analogous calculation for D mesons, 
$g_{D^*D\pi}= 12.5\pm 1$. 
Since these predictions depend on light-cone 
wave functions at a fixed point, they are less certain
than the results on form factors which involve integrals  
over wave functions.

With the above value of $g_{D^*D\pi}$ one obtains the hadronic width
$\Gamma (D^{*+} \ra D^0 \pi^+ )=32 \pm 5~ keV$.
This estimate is perfectly consistent with the upper limit derived from
recent measurements \cite{ACCMOR,CLEO1}.

\section{ Extraction of $V_{ub}$}

Furthermore, from the results on $f^+(p^2)$ and $F_0$ shown in Fig. 1,  
one can compute the widths for
$B^0 \ra \pi^- l^+ \nu_l$ ( $ l=e,\mu,\tau$).
We find
\be
\Gamma(B^0\rightarrow\pi^- e^+ \nu_e) = 
8.7 ~|V_{ub}|^2~\mbox{ps}^{-1}~,
\label{Bwidth}
\ee
\be
\Gamma(B^0\rightarrow\pi^- \tau^+ \nu_{\tau}) = 
(0.7 \div 0.8) \Gamma(B^0\rightarrow\pi^- e^+ \nu_e). 
\label{Bwidth1}
\ee
Unlike $\Gamma(B^0\rightarrow\pi^- e^+ \nu_{e})$, the 
width of $B^0\rightarrow\pi^- \tau^+ \nu_{\tau}$ is 
very sensitive to the form factor $F_0$.
The interval given in (\ref{Bwidth1}) 
reflects the uncertainty in the extrapolation of $F_0$ 
from the endpoint of the curve in Fig. 1 to the value 
$f_B/f_\pi = 1.0\div 1.7$ at $p^2= m_B^2$  
implied by current algebra \cite{Vol}. 

Experimentally, combining the recent 
CLEO result for $l=e,\mu$ \cite{CLEOBpi}, 
$BR(B^0\ra\pi^-l^+\nu_l) = (1.8 \pm 0.4\pm 0.3 \pm 0.2)\cdot 10 ^{-4}$~,
with the world average of the $B^0$ lifetime
\cite{PDG},  $\tau_{B^0}=1.56 \pm 0.06 $ ps,
one obtains
$
\Gamma(B^0\rightarrow \pi ^- l^+ \nu_l) = 
(1.15 \pm 0.34)\cdot10^{-4}~\mbox{ps}^{-1}~,
$
where the errors have been added in quadrature.
Comparison with (\ref{Bwidth}) then yields
$ 
|V_{ub}| = 0.0036 \pm 0.0005~.
$
The additional theoretical uncertainty is estimated
to be less than 20\%.

Recently, also the $B\ra \rho$ form factor 
has been calculated with the help of a light-cone sum rule \cite{Ball}.
The resulting width 
$\Gamma(B^0\rightarrow \rho ^- l^+ \nu_l)= (14\pm 4) 
~|V_{ub}|^2~\mbox{ps}^{-1}~$ together with (\ref{Bwidth}) 
yields a $\rho/\pi$ ratio in agreement with
the CLEO finding \cite{CLEOBpi} of about 1.4 .

\section{Radiative decays} 

A further class of processes to which the method of 
light-cone sum rules can be applied fruitfully are 
exclusive radiative $B$ and $D$ decays. Recent examples
include the calculations of the magnetic penguin 
form factor \cite{ABS} in $B\rightarrow K^*\gamma$ and $\rho \gamma$, 
and of the contribution from weak annihilation 
\cite{KSW,AB} to $B \to \rho \gamma $.  
In the latter case, 
the light-cone expansion leads to matrix elements,
respectively, hadronic wave functions associated with 
photon emission by a quark-antiquark pair
at light-like separation. From these estimates  
weak annihilation is expected to contribute 
to $B  \rightarrow \rho \gamma $ at most $10 \%$
of the penguin mediated short-distance amplitude.
This effect 
is comparable in size to the theoretical uncertainty in the main 
amplitude. However, 
with increasing precision
the long-distance effect due to
weak annihilation  may become non-negligible.

In radiative $D$ decays, weak annihilation plays a more
pronounced role since the short-distance penguin contributions are
completely negligible.
Light-cone sum rules \cite{KSW} predict the branching
ratios $BR(D^{0} \rightarrow \bar K^{*0}\gamma)=
1.5 \cdot 10^{-4}~$ and
$BR(D_s \rightarrow \rho^+ \gamma = 2.8\cdot 10^{-5}$.
The experimental observation of these modes is an interesting task for 
the next generation of charm experiments.  

Finally, by replacing the light hadronic currents in 
the above applications by a leptonic current, one can treat decays 
like $B\ra l \nu_l \gamma$ in same framework. 
These modes are very interesting as the strong 
suppression for $l= e,\mu$ due to helicity conservation should 
be lifted by the photon emission. Indeed, explicit
calculation \cite{KSW} predicts a factor 10 enhancement:
$\Gamma(B\rightarrow \mu\nu_\mu\gamma)/
\Gamma(B\rightarrow \mu \nu_\mu )
\simeq 11.5(180 \mbox{MeV}/f_B)^2$.

With the few examples selected for this review we hope to have
illustrated the usefulness of QCD sum rules on the light-cone. 
\bigskip
\section*{Acknowledgments}
A.K. is grateful to G. Stoll and D. Wyler for 
useful discussions on common results presented here.
This work is supported by the Bundesministerium 
f\"ur Bildung und Forschung 
(BMBF) under contract 05 7WZ91P(0), and by the 
EC program Human Capital and Mobility (HCM) under contract CHRX-CT93-0132.

\newpage

\begin{figure}[htbp]
\begin{center}
\vspace{6.5cm}
\hspace{-8.0cm}
\mbox{
\epsfysize=7cm
\epsffile[0 0 300 300]{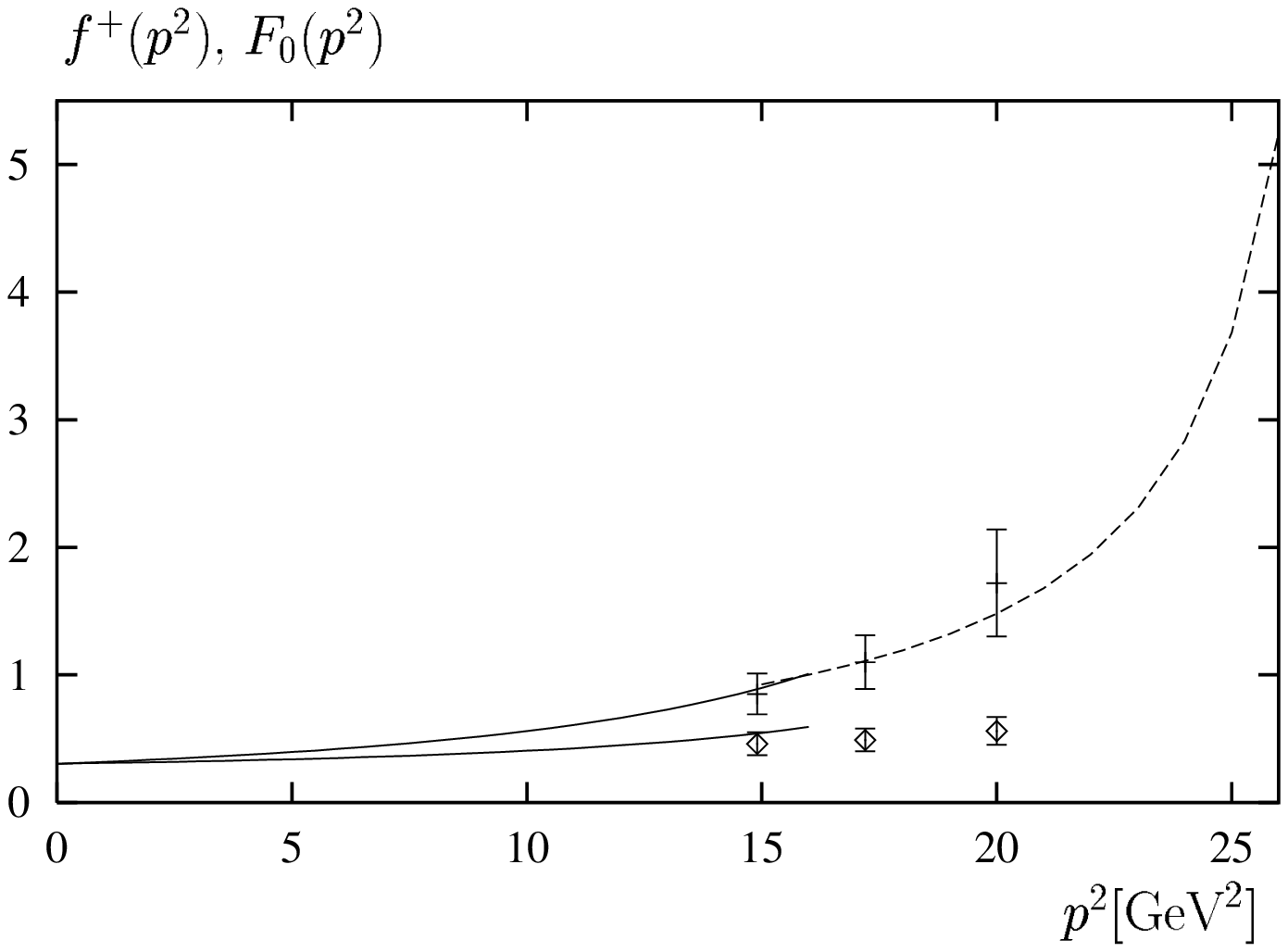}}
\end{center}
\vspace{-6.8cm}
\caption[]
{$B \rightarrow \pi$ form factors: 
$f^+$ (upper curve extrapolated 
using single-pole approximation)
and $F_0$ (lower curve). The data points indicate
lattice results \cite{lat}. 
\label{f2}
}
\end{figure}

\vspace{2cm}

\begin{figure}[htbp]
\begin{center}
\vspace{5.0cm}
\hspace{-8.0cm}
\mbox{
\epsfysize=7cm
\epsffile[0 0 300 300]{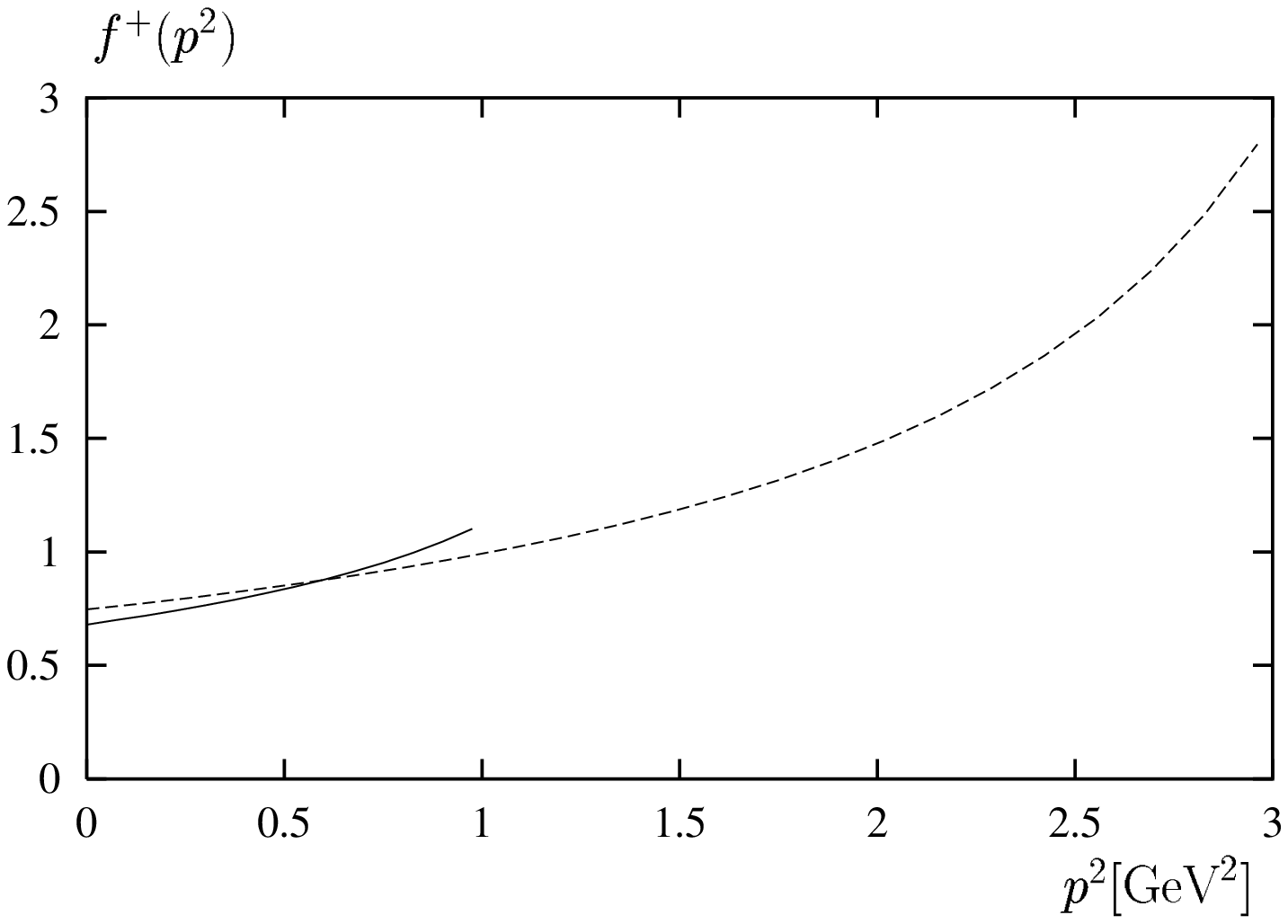}}
\end{center}
\vspace{-6.8cm}
\caption[]
{ The $D \rightarrow \pi$ form factor $f^+$: 
direct sum rule result (solid curve) 
in comparison to the single-pole approximation
(dashed curve). 
\label{f3}
}
\end{figure}

\end{document}